\documentclass[preprint]{aastex}
\shorttitle{Optical Survey of Unidentified EGRET Sources}
\shortauthors{Bloom et al.}
\begin{document}
\title{An Optical Survey of the Position Error Contours of Unidentfied High Energy Gammma-ray   Sources at
Galactic Latitude b$ > \vert 20^{\circ} \vert $}
\author{S. D. Bloom \altaffilmark{1}, D. A. Dale \altaffilmark{2}, R. Cool \altaffilmark{2,3}, K. Dupczak \altaffilmark{2}, C. Miller \altaffilmark{2}, A. Haugsjaa \altaffilmark{2}, C. Peters \altaffilmark{1}, M. Tornikoski \altaffilmark{4}, P. Wallace \altaffilmark{5}, M. Pierce \altaffilmark{2}} 
\altaffiltext{1}{Department of Physics \& Astronomy, Hampden-Sydney College, Box 821, Hampden-Sydney, VA 23943}
\altaffiltext{2}{Department of Physics \& Astronomy, University of Wyoming, Laramie, Wyoming 82070}
\altaffiltext{3}{Steward Observatory, University of Arizona, Tucson, AZ 85721}
\altaffiltext{4}{Metasahovi Radio Observatory}
\altaffiltext{5}{Department of Physics \& Astronomy, Berry College, Mount Berry, GA 30149} 
\begin{abstract}
We present results from our optical survey of the position error 
contours (``error boxes'')
of unidentified EGRET sources at mid to high Galactic latitude. It is our 
intention to search for potential blazars that may have been missed in the 
original identification process of the three EGRET Catalogues and supplementary publications. We have first 
searched
the error contours of unidentified sources at $b > \vert 20^\circ \vert$ for 
flat spectrum radio sources using the NASA Extragalactic Database (NED). For 
each such radio source found we conducted optical searches for counterparts 
using the Palomar 60-inch telescope, and University of Wyoming's 2.3 and 0.6 m telescopes. Many of the 
radio sources have plausible optical counterparts, and spectroscopy will be 
conducted at a later date to determine which of these sources are quasars or 
active galaxies. Results show thats several sources are optically variable, and/or have flat or inverted radio to millimeter spectra and are thus potential blazars.
 
\end{abstract}


\section{Introduction}
Though the EGRET instrument detected 271 sources, most of these sources 
remain unidentified with counterparts at any other wavelength \citep{hartman99}. However, 
the distribution of these sources on the sky does suggest at least two major 
populations: Those near the Galactic plane, where 90 \% of the sources are 
unidentified, and those above the plane, where 50 \% are unidentified \citep{caraveo01}.
\citet{caraveo01} further suggest that those sources above the plane may have at least 
three subcomponents: extragalactic, Gould Belt, and Galactic halo. Though 
nearly all of the firmly identified extragalactic sources are blazars, there 
have been some recent studies suggesting that some sources are related to 
non-blazar AGN \citep{mukherjee02} or galaxy clusters \citep{colafrancesco02}. The task of investigating all of these sources seems 
daunting. Therefore, several investigators have gone the route of examining 
several individual sources in depth. An advantage of this approach is that 
such searches are like to yield abundant information on the several sources 
they investigate; however, it is more difficult to draw general conclusions from such results. We continue along 
these lines by searching within the error contours of high Galactic latitude 
EGRET unidentified sources for new blazars. We will then ascertain whether 
any such objects found are likely to be the source of the gamma rays.
At the time of submission, we became aware of a parallel study by
\citet{sowards03}. Though similar in scope, ours studies differ, and are complimentary, in that we are continuing multi-epoch optical observations with B, V, R photometry and do include some sources at southern declinations.
\section{Source Selection}
Using the NASA Extragalactic Database (NED), we have searched for flat
 spectrum radio sources (spectral index,$ -0.5 < \alpha < 0.5$) within the 
95 \% positional confidence contours of EGRET unidentified sources with 
$ \vert b \vert > 20^\circ$. Unless otherwise mentioned, spectral indices referred to in this paper are between 1.4 and 5 GHz. Aside from a handful of pulsars and relatively nearby extragalactic sources, blazars with flat radio spectra are the only objects that have been detected by EGRET as a class. Though not all flat spectrum sources are blazars, it remains a good way of separating out 
possible blazars from a much larger group of radio sources. We have chosen 
this range of Galactic latitudes so that there will be minimal overlap with 
Galactic populations discussed in the introduction and fewer problems with optical identification in crowded fields. It should be noted, however, that EGRET blazar candidates have been found in the Galactic plane \citep{halpern01}.The radio data referenced in NED were usually from the  Green Bank Surveys, which had positional acccuracies of about 1-3 arcminutes (depending on the observing 
frequency and the source brightness). This resolution is far too coarse for conducting 
optical searches. We have therefore additionally searched for these sources in 
the NRAO VLA Sky Survey (NVSS) in order to obtain positions accurate to 
approximately 1 arcsecond, which can be achieved for sources brighter than
15 mJy \citep{condon98}. 
In addition to the flat spectrum radio 
sources, we have compiled a secondary observing list of radio sources of 
unknown spectral index, and or X-ray and optical sources with accurate 
positions. 

After applying these criteria, 53 EGRET sources remain in the potential sample (unidentified in the Third EGRET Catalogue), with an average of about 1 candidate radio source within the position error contours of the EGRET source. In addition, there are 18 AGN's with questionable identifications. We further 
restricted ourselves to the ranges of right ascension and declination that were visible 
from Mount Palomar during the nights in November 1999 and from the Wyoming 
telescopes during the subsequent nights in 2002.  These limits lead us to exclude all sources with declination below -30 $^\circ$ , leaving us with 41 unidentified sources and 13 with questionable identifications. The right ascension limits are somewhat less strictly determined since we observed at several different points during the year. However, Northern Hemisphere summer months were those most deliberately avoided (usually due to poor sky conditions). Sources in the 18h to 0h range would be least covered by the observations (about half of the remaining sources). 

An extensive study of the identification of radio sources with EGRET sources 
was conducted by \citet{mattox97,mattox01}. Essentially they conclude that it is difficult to 
positively identify a gamma-ray source with any radio source with flux 
density less than about 500 mJy at 5 GHz mostly because the sky density of 
such sources is so high that the probability for chance correlations is 
correspondingly high. They assume that the radio spectra would be near their 
peak fluxes at these frequencies, and thus that the 5 GHz flux densities 
would be representative of the peak synchrotron output of the source. This 
assumption was likely partially made out of necessity, since there are no complete high 
frequency surveys with sensitivity less than 1 Jy. However, other investigators have 
shown that several sources might be fairly dim at these low frequencies and 
have a brighter (and flatter, more variable)
spectrum extending to at least 200 GHz
\citep{tornikoski02,bloom97,bloom00}. Thus some dim 5 GHz sources, which by
 the method of \citet{mattox97,mattox01} would not be considered to be counterparts to 
EGRET 
sources may indeed be counterparts to gamma-ray blazars. We intend to make up 
for this deficiency by searching for optical counterparts to radio sources that may 
be dim (in some cases, $<$ 100 mJy) at 5 GHz. If a counterpart is found 
we will search for blazar-like activity, and go on to investigate the optical 
and radio spectra of these objects. 
\section{Observations}
We began conducting the search for optical counterparts on 3-4 November 1999 
using the 60-inch telescope at Mount Palomar with a Tek 2048 x 2048 CCD. The 
pixel size in this set-up is 0.37 arc seconds; however, we have  binned 
the pixels 2-by-2 to an effective 
size of 0.74 arc seconds. Since the seeing for those nights (and most at 
Palomar) was worse than 1 arc second, no crucial information on source 
structure was lost. In addition, we would expect most of our sources to be 
point-like. We have used the Johnson B, V, and Kron R filters for determining 
magnitudes in those bandpassses. The procedure was to first observe a 
candidate source in R for 10 minutes (and then adjust to shorter or longer 
exposures depending on the results). If no obvious counterpart was found after
 20 minutes, we did not go to the other filters. If a potential source was 
seen, we observed in V and B as well.
In addition to the program sources, we have observed Landolt standard stars 
throughout the evening to determine the photometric scale and atmospheric 
absorption \citep{landolt92}. 

At Red Buttes (RBO) imaging was completed using an SBIG ST-6 CCD Camera.
The ST-6 was mounted at Cassegrain focus, as is the common user mode at RBO. 
The pixels were 27 $\mu$m by 23 $\mu$m with a field of view of 4.7 arcminutes x 6.2 arcminutes. The plate scale is 1.14 arcseconds x 0.97 arcseconds.
Standard filters, as described above, were used. Again, standard stars were used to calibrate source magnitudes \citep{landolt92}.

At the Wyoming Infrared Observatory (WIRO), an Apogee-AP8 camera was used on the 2.3 m.
A new configuration, WIRO-Prime, was used in which camera was mounted at prime focus, giving a 18 arcminutes x 18 arcminutes field of view \citep{pierce02}. The pixel size of 24 x 24 $\mu m
^2$ gave a plate scale of 1 arc second per pixel. The seeing in May 5, 2002
ranged between 2-3 arcseconds throughout the night. Filters identical to those used at RBO were used for multicolor photometry.

Data reduction was carried out uisng the standard photometry packages in the
Image Reduction and Analysis Facility (IRAF). The atmospheric extinction values for airmass and color corrections were calculated using the STELLARPHOT package developed and maintained by Cornell University.

In addition, some sources with high enough 5 GHz flux were additionally observed at 37 GHz, at the Metsahovi Observatory in Finland, and at 90 GHz using SEST in Chile (see \citet{tornikoski02} for more information on the observing procedure at those telescopes). These results are discussed below in Section 5.

J. Halpern (Columbia University) has provided us with the optical spectrum of one source, NVSS J032850+212825. The data were acquired at the KPNO 2.1 m on 23 Oct 2001 using the Goldcam spectrograph and Ford 3K x 1K pixel CCD with 15 micron pixels. The spectrograph uses a grating of 500 l/mm, blaze 550 angstroms and resolution of 5 angstroms. The slit width is 1.9 arcseconds. More details of the spectographic set up can be obtained at http://www.noao.edu/kpno/manuals/l2mspect/spectroscopy.html.
The significance of this particular spectrum is discussed below.
\section{Analysis}
Our first goal in the analysis was to locate potential matches between the NVSS radio sources and an optical source. Clearly, this could
only be achieved once an absolute coordinate system was established for the images (discussed below). The approximate radio source position accuracy is given by:

$$ \sigma_p= {0.5 \theta {\sigma_s } \over S}$$ 
where $\sigma_p$ is the position uncertainty, $\theta$ is the Gaussian beam size in arc seconds, $\sigma_s$ is the uncertainty of the flux and $S$ is the 1.4 GHz flux from the NVSS. If we subsitute in approximate numbers, $\theta=40$, ${\sigma_s /S}=25$ , then $\sigma_p=0.8$.
We thus have to acheive accuracy somewhat better than this to be certain of 
an optical coincidence with a radio source. Practically speaking, if the offset between the NVSS position and optical position is much greater than this uncertainy (typically 0.5-1.0 arc seconds for our sample), we can not be very confident in our match.
 We have matched the optical images to the coordinate scale of the USNO-A2.0 catalogue using the XTRAN routine within AIPS. Tables 1-3 show the results for each telescope.  Most of the radio sources have plausible visual counterparts; however some have large uncertainty of coincidence between the radio and optical source. Below, we consider the most likely matches to the gamma ray source (if any). We have also quantified the amplitude of the optical variability with the parameter:
$$ A \equiv {{F_{max}-F_{min}} \over {F_{max}+ F_{min}}}$$
Where the $F$ values correspond to the minimum or maximum flux at a particular waveband. A factor of 2 in variabilty ccorresponds to $A=0.333$.
Though we also include the USNO magnitudes for comparison in the individual source notes below, we note that this is primarily for qualitative comparison to the current measurements. In particular, \citet{monet03} estimate an overall uncertainty in photometry of 0.25 mag in the USNO catalog. However, the photometry of extended sources is considerably worse, and even that of stars can be worse if there are no photometric standard stars on or overlapping the plate of interest. Since many of our identifications are extended sources, we cautiously conclude that even marginal variability (when using USNO catalog magnitudes) can not be suggested unless $\Delta mag \sim 1$ or greater. We have also included information on the known gamma-ray variability of the EGRET sources, in the form of the $\delta$ statistic of \citet{nolan03}. This statistic divides the standard deviation of the flux by the average for a specific gamma-ray source. For each source, a most likely value of $\delta$, and a 68\% confidence interval are determined assuming a gamma distribution for the fluxes. They also calculate a probability that the source shows variability consistent with being beyond that caused by systematic uncertainty alone . They define a statistic, $V_{12}$, described by the following equation, that allows for a less cumbersome representation of the probabilities:
$$ {V_{12}}=-log_{10}[1-P_{\chi^2}(r|1)]$$

Here, $P_{\chi^2}(r|1)$ refers to the cummulative $\chi^2$ probability distribution with one degree of freedom. See \citet{nolan03} for justification of using this probability distribution.
With these parameters, $\delta > 1$ with $V_{12} >1$ are likely to be truely variable. To contextualize these numbers more clearly, 3C 279, which is widely considered as a highly variable source, has $\delta=0.9$ and $V_{12}=\infty$. The Crab Pulsar, a steady source, has $\delta=0.08$, and $V_{12}$ is undetermined because $\delta$ is below that expected from systematic uncertainty alone. $\delta =0$ or near 0 implies non-variability within the systematic uncertainty of gamma-ray fluxes.$V_{12}=\infty$ implies a 0 (or nearly) probability of the source being constant.
We disucss this analysis below in the context of possible multiwavelength variability of each source. 

\section{Notes on Individual Sources}
3EG J0038-0949.- NVSS J003906-094247 is the most plausible counterpart for this source, and is the only possible blazar counterpart discussed in the literature thus far \citep{mattox01}. The optical and NVSS positions are offset by 0.83 arcseconds (approximately same as NVSS position uncertainty). \citet{reimer03} consider the possibility that this source is associated with the A85 galaxy cluster, however conclude that it is likely to be a chance coincidence. The flux at 5 GHz is approximately 0.2 Jy , $\alpha=
-0.1$. This source exhibits marginal night to night optical variability, as indicated in Table 1. We also note the possible long term variability, particlarly in B, as determined from the comparison to the USNO magnitudes from 1954 (B=19.4; R=19.1). 
The other possible identification we consider (PMN J0037-1010; NVSS J003750-101034) is a much weaker (0.04 Jy at 5 GHz) and steeper spectrum radio source ($\alpha=-0.7$). Additionally, this source shows no evidence for optical variability. In this case, the optical and NVSS sources are offset by 2.5 arcseconds, and thus the radio/optical match is somewhat questionable.
The $\gamma$-ray variability ($\delta=0$) is insignificant .  
\vskip 1mm
3EG J0215+1123.- We consider two potential identifications for this object. Though the QSO B0214+108 (4C 10.06) is a moderately strong Green Bank Survey source at 5 GHz (0.44 Jy), the two nearest (and apparently related) compact NVSS object are significantly offset from the optical source (45 arc seconds). The spectral index is -0.8, steeper than expected for a typical blazar. However, this source has been detected recently at 37 GHz with a flux of 0.3 Jy. This source was not detected at 90 GHz (implying a flux $<$0.4 Jy).
This is the only possible identification mentioned by \citet{mattox01}. The complex radio structure of 4C 10.06 has been likened to relatively nearby wide-tailed radio galaxies
whose radio structure is created by interaction with an intergalactic medium.
In this particular case, the QSO may be related to a cluster of galaxies \citep{hintzen84}. This would open up the possibility that at least some of the $\gamma$-ray  emission (and X-ray flux as well) is not directly related to the AGN \citep{nolan03}.

The other possible identification, NVSS J021527+112318, has a flux of only 0.132 Jy at 5 GHz, but also has been detected at 0.3 Jy level at 37 GHz, thus indicating a rising, and/or variable spectrum. This source was not detected at 90 GHz (flux $<0.4$ Jy). The optical and NVSS positions are offset by 4.17 arcseconds and thus somewhat suspect (a few times the uncertainty). If, however, 4C 10.06 is the source of the $\gamma$-rays, there may be a high frequency blazar compenent to that radio source that isn't apparent in the surveys, that only show 
the extended low frequency source. Night to night variability for the QSO related to 4C 10.06 (in R only) is evident at the $4-\sigma$ level.
The USNO magnitudes do not strongly suggest long term variability (B=18.6, R=16.9). 
Note that these candidates are in addition to the identification favored by \citet{sowards03}, which we have not observed. Gamma-ray variability is observed ($\delta=1.27 $, with $V_{12}=2.10$).
\vskip 1mm
3EG J0245+1758.-The most plausible identification is NVSS J024640+180144. This radio source has flux of about 0.2 at 5 GHz and $\alpha=0.1$. \citet{sowards03} list this as a potential identification. No BVR night to night variability is seen; however, a possibility of long term variability is suggested by the USNO magnitudes. The optical and NVSS positions ar offset by 1.03 arcseconds (comparable to the uncertainties in position). A second potential identification would be NVSS J024437+17221, however this source is very dim at 5 GHz (0.033 Jy ;$\alpha=-0.1$). This source also shows no evidence for optical night to night variability or long term variability, considering both the USNO magnitudes (B=19.6; R=18.9) and recent measurements with RBO. The $\gamma$-ray variability is marginal ($\delta=1.14$; $V_{12}=1.38$).
\vskip 1mm
3EG J0329+2149.-We find that NVSS J032850+212825 is a potential identification.Though this is a dim radio source at 5 GHz (0.057 Jy), we find it additionally interesting because it is an X-ray (ROSAT) and infrared (2MASS) source as well. \citet{punsly99} argue that this source, among others, may be an intermediately peaked BL Lac object (IBL), producing X-rays in a relativistic jet. Such sources would be expected to have relatively large X-ray fluxes even if the radio source is weak. As explained in \citet{punsly99} and references therein, an IBL could have relatively strong magnetic fields, leading to relatively low radio flux, but relatively higher optical through X-ray flux as compared to blazars that have spectral peaks at lower frequencies (LBL) It was not detected at 37 GHz during recent observations at Metsahovi, implying a flux below 0.2 Jy. This source was also not detected at 90 GHz, implying a flux below 0.4 Jy.  The optical and NVSS positions are offset by 3.59 arcseconds ( a few times the uncertainty).
A recent optical spectrum (Figure 1) shows evidence for a redshifted Ca II H \& K feature at 5200 angstroms (z=0.31), though the feature is weak. Imaging shows that this object is extended, and thus is likely at  z $<$ 0.3 \citep{halpern03}. We therefore consider the redshift to plausible for this object. In addition to the extended appearance of the objects, the infrared colors of 2MASS suggest the source is non-stellar \citep{tokunaga00}. \citet{sowards03} suggest a different source for the identification. No short term optical variability is indicated in our data, nor is there any significant difference in our magnitudes and those of the USNO catalogue. The source is not variable in gamma rays ($\delta=0$).
\vskip 1mm
3EG J0622-1139.-The radio source has already been reported as a possible identification \citep{bloom00}, but we present a new epoch of visual data. These data, and the colors of 2MASS suggest the source is non-stellar \citep{tokunaga00}. This source is marginally variable in gamma rays ($\delta=1.04$; $V_{12}=1.08$)
\vskip 1mm
3EG 0852-1216.- PMN J0850-1213 was reported as an identification by \citet{halpern97}, and subsequently seen to be highly variable in optical and radio/millimeter \citep{halpern97,bloom97}. The optical and NVSS sources are offset by 0.16 arcseconds (much less than the uncertainties in position). We present a new epoch of visual data here, and conclude that the object was in a low optical state at the time of these observations. The history of this object points out that some ordinary Green Bank survey sources can actually be distinguished blazars in disguise. Variability is indicated by the $\delta=1.21 $ and associated probability ($V_{12}=1.94$).
\vskip 1mm
3EG J1052+5718.- The possible identification with NVSS J105837+562811 was reported previously \citep{bloom97}, and we include new optical photometry here. The optical source appears to be significantly dimmer than in the POSS plates (B=14.7, R=14.2). Similar to 3EG J0329+2149 discssed above, this source is also considered to be an IBL by \cite{punsly99}. This is source shows no $\gamma$-ray variability  ($\delta=0.24$; $V_{12}=0.10$). The optical and NVSS sources are offset by 0.89 arcseconds (approximately same as uncertainties in position). 
\vskip 1mm
3EG J1236+0457.- There is a possible identification with B 1237+0404 (NVSS J123934+035020), previously mentioned in \citet{mattox01}. We present multicolor photometry for the first time. There are no indications of short or long term optical variability. There is no evidence of $\gamma$-ray variability ($\delta=0.53$; $V_{12}=0.10$).
The optical and NVSS sources are offset by 1.17 arcseconds (approximately same as uncertainties in position).
\vskip 1mm
3EG J1337+5029.- There are no radio sources in this field $>0.045$ mJy, however we report on the optical observations of two X-ray sources in the field. These two X-ray sources, and the others in this region appear to be associated with normal galaxies. 
If the gamma-ray source is related to either of these X-ray sources, but has no radio counterpart, then it must either not be an AGN, or of a type not previously known to have been detected by EGRET. With no further information, we do not consider these sources to be compelling counterparts. The source is not variable in $\gamma$ rays ($\delta=0.53$; $V_{12}=0.41$).
\vskip 1mm
3EG J1600-0351.- A possible identifcation is PMN J1601-0302 (NVSS J160117-030231), with
5GHz flux of 0.06 Jy. This source is coincident with an extended visual (POSS and our survey) and infrared object (2MASS). The radio source is dim at 5 GHz, and is thus not a likely counterpart, based soley on that information. The optical and NVSS sources are offset by 6.21 arcseconds (10 times larger than the uncertainty in position). However, this source should be monitored across the broad band spectrum for signs of blazar activity. This source is not gamma-ray variable ($\delta=0.01$)  
\section{Conclusions}
We have searched the ``error boxes'' of EGRET sources, and found several potential blazar counterparts at other wavebands. The most compelling case, newly reported here, is that of 3EG J0038-0949 (matched with NVSS J993906-094247). These counterparts are generally characterized by flat or inverted radio spectra (in some cases up to 37 GHz), and in some cases, long and short term optical variability. We conclude that several of these objects are the source of the gamma rays. This study and others have shown that the EGRET instrument detected a class of dim blazars, in addition to those identified in the Third EGRET Cataloge (\citep{hartman99}. NASA's upcomming Gamma Ray Large Area Telescope (GLAST) mission should confirm the existence of a relatively dim population of blazars thatare the source of gamma rays for many of the unidentified sources at high Galactic latitude.
\acknowledgements
S. Bloom acknowledges an American Astronomical Society Small Research Grant, a Mednick Fellowship from the Virginia Foundation of Independent Colleges and several Summer Faculty Fellowship grants from Hampden-Sydney College. R. Cool and K. Dupczak were supported by Wyoming NSF/EPSCoR. A. Haugsjaa and C. Miller  are supported by NSF REU site grant AST0097356.
This research has also made use of the NASA Extragalactic Database.
The Digitized Sky Surveys were produced at the Space Telescope Science Institute
under a U. S. Government grant NAG W-2166. The images of these surveys
were based on the photographic data obtained using the Oschin Schmidt 
Telescope on Palomar Mountain and the UK Schmidt Telescope. The plates
were processed into the present compressed digital form with the permission of 
these institutions. The Guide Star Catalogue was also produced at the Space
Telescope Institute using the same photographic database.
STELLARPHOT is part of the GALPHOT surface photometry package, a collection of IRAF/SDAS scripts originally developed by W. Freudling and J. J. Salzer and maintained at Cornell University by M. P.Haynes. STELLARPHOT itself has been significantly developed by L. E. Van Zee.  
{}
\newpage
\centerline{\bf FIGURE CAPTIONS}
\figcaption[figure1.ps]{Optical spectrum of NVSS J032850+212825 using KPNO 2.1 m and Goldcam Spectrograph}

\end{document}